\documentclass[journal,twoside]{IEEEtran}

\usepackage[cmex10]{amsmath}      
\usepackage{amssymb,amsfonts}
\usepackage{amsthm}
\usepackage{bm}

\usepackage{cite}

\usepackage{graphicx}
\usepackage[caption=false,font=footnotesize]{subfig} 
\usepackage{booktabs}
\usepackage{multirow}
\usepackage{array}
\usepackage{tabularx}
\usepackage{float}

\usepackage{algorithm}
\usepackage[noend]{algpseudocode}

\algnewcommand{\Inputs}[1]{%
  \State \textbf{Inputs:}
  \Statex \hspace*{\algorithmicindent}\parbox[t]{.8\linewidth}{\raggedright #1}
}
\algnewcommand{\Initialize}[1]{%
  \State \textbf{Initialize:}
  \Statex \hspace*{\algorithmicindent}\parbox[t]{.8\linewidth}{\raggedright #1}
}

\usepackage{xfrac}
\usepackage{textcomp}
\usepackage{listings}
\usepackage{soul}
\usepackage{tikz}
\usepackage{stackengine}
\usepackage{cancel}
\usepackage{flushend}
\usepackage{hyperref}
\usepackage{amsmath}
\allowdisplaybreaks[3] 
\usepackage{xcolor}
\usepackage{placeins}

\hypersetup{
     colorlinks   = true,
     linkcolor    = blue,
     citecolor    = black,
     urlcolor     = blue
}
\usepackage{breqn}

\makeatletter
\newcommand*{\transpose}{%
  {\mathpalette\@transpose{}}%
}
\newcommand*{\@transpose}[2]{%
  \raisebox{\depth}{$\m@th#1\intercal$}%
}
\makeatother

\usepackage{graphicx}
\usepackage{dblfloatfix} 

\newtheorem{remark}{Remark}

\begin{document}
\renewcommand{\ttdefault}{cmtt}
\bstctlcite{IEEEexample:BSTcontrol}

\title{Scalable Quantum Clustering of Renewable Profiles via a Hybrid Qubit-efficient Strategy}



\author{
{Yifeng Cao,~\IEEEmembership{Student Member},
Chengjun Liu,~\IEEEmembership{Student Member},
Yijun~Xu,~\IEEEmembership{Senior Member}, 
Kai Wen,
Wei Gu,~\IEEEmembership{Senior Member},
Lamine Mili,~\IEEEmembership{Life Fellow},
and
Shuai Lu

}
\thanks{Y. Cao, Y. Xu, W. Gu, C. Liu and S. Lu are with the  Electrical Engineering Department, Southeast University, Nanjing, Jiangsu 210096, China, (e-mail: \{caoyf, yijunxu, wgu, liucj, shuai\_lu\}@seu.edu.cn).}
\thanks{K. Wen is with Beijing QBoson Quantum Technology Co., Ltd., Chaoyang District, Beijing 100020, China (e-mail: wenk@boseq.com).}
\thanks{L. Mili is with the Bradley Department of Electrical and Computer Engineering, Virginia Tech, Northern Virginia Center, Falls Church, VA 22043 USA (e-mail: lmili@vt.edu).}
}
\maketitle
\markboth{‌CSEE JOURNAL OF POWER AND ENERGY SYSTEMS‌}%
{Cao \MakeLowercase{\textit{\textit{et al.}}}: Scalable Quantum Clustering of Renewable Profiles via a Hybrid Qubit-efficient Strategy }

\begin{abstract}
Large-scale renewable energy profile clustering is a computationally challenging NP-hard problem. Recent quantum computing techniques offer a potential remedy, yet their scalability is severely constrained by the limited qubit count of current quantum hardware.  To address this issue, we first cast the problem into a QUBO form and then we propose a consensus-based variable reduction (CBVR) method to drastically cut qubit usage. More specifically, we tailor the one-hot constraint to eliminate assignment variables and leverage consensus analysis to convert stable clustering relations from initial anneals into equality constraints for further reduction. Experiments on real quantum hardware confirm that CBVR compresses oversized QUBO models into feasible sizes while retaining high clustering quality and efficiency.
\end{abstract}

\begin{IEEEkeywords}
Renewable energy profile, Clustering, co-assignment analysis. 
\end{IEEEkeywords}
\IEEEpeerreviewmaketitle
\section{Introduction}
\IEEEPARstart{C}{lustering} is widely used in power systems for applications such as data analysis and aggregation control \cite{jin2022wind,amini2022effective}. It can be formulated as a constrained combinatorial optimization problem that is NP-hard in general \cite{xu2005survey}. 

However, as renewable energy sources become increasingly integrated, the computational difficulty of solving the clustering problem increases rapidly. To address this challenge, classical methods such as RSPCA
\cite{mahmud2025rspca}, RSPCE \cite{mahmud2023rspce}, and
ACEM \cite{mahmud2023acem} have achieved effective and scalable
clustering through random sampling and ensemble learning. Nevertheless, the complexity of large-scale clustering, together with the finite processing and memory resources of classical computers, can still make finding a global optimum computationally demanding \cite{liu2026alleviating,mahmud2025rspca}. This motivates the exploration of more efficient methods for large-scale renewable energy profile clustering.

Fortunately, quantum computing has been explored for clustering through algorithms such as quantum K-means \cite{DiAdamo2022} and quantum K-medoids \cite{PhysRevA.107.022421}. These studies demonstrate the potential of quantum computing for renewable energy profile clustering, although its practical implementation is limited by the qubit capacity of current quantum hardware.

Since clustering problems are often formulated as QUBO models for quantum optimization, reducing the size of QUBO models has become an important strategy for reducing qubit requirements. To address this limitation, recent studies have proposed methods such as subQUBO \cite{Atobe2022} and spin-variable reduction \cite{shirai2023spin}. However, the former suffers from unstable solution accuracy due to the inherent randomness in the subset selection, while the latter's performance is heavily limited by the number of equality constraints.

To address this issue, this paper proposes the consensus-based variable reduction (CBVR) method, which exploits stable co-assignment  relations to generate additional equality constraints for clustering renewable energy profiles on QUBO-based quantum hardware, such as coherent Ising machines (CIMs). We first formulate the renewable energy profile clustering problem as a QUBO model based on Euclidean distance and one-hot constraints. Then, the one-hot constraints are reformulated to eliminate one assignment variable for each profile. Moreover, stable co-assignment  relations identified from preliminary annealing runs are further converted into additional equality constraints for variable reduction. Experiments on real quantum hardware verify the effectiveness of the proposed method in improving the scalability of quantum clustering.

\section{Preliminaries}
\subsection{Renewable Energy Profile Clustering Model}
A renewable energy profile represents the power output recorded at each time point of the renewable energy stations throughout the day. In this paper, wind turbine profiles are used as a representative example. Suppose that \(N\) profiles are to be clustered into \(G\) groups. The objective of clustering is to minimize the intra-group dissimilarity. 

Then, let \(x_i^g \in \{0,1\}\) be a binary decision variable, where \(x_i^g=1\) indicates that renewable profile \(i\) belongs to Group \(g\); otherwise \(x_i^g=0\). Accordingly, the objective function, $H_{\mathrm{obj}}$, is
\begin{equation}
\min \; H_{\mathrm{obj}}=
\sum_{i=1}^{N}\sum_{j=1}^{N} w_{i,j}
\sum_{g=1}^{G} x_i^g x_j^g ,
\label{eq:clustering_obj}
\end{equation}
subject to the one-hot constraints given by
\begin{equation}
\sum_{g=1}^{G} x_i^g = 1,
\qquad \forall i \in \{1,2,\dots,N\},
\label{eq:assignment_constraint}
\end{equation}
where \(w_{i,j}\) denotes the Euclidean distance between the profiles \(i\) and \(j\).
This formulation is a binary constrained combinatorial optimization problem. To make it compatible with quantum hardware, we need to reformulate it as a QUBO model \cite{liu2026alleviating,liu2026quantum}.

\subsection{The Ising Model}
The Ising model translates combinatorial optimization into finding the ground state of a system of interacting spins, where each spin configuration encodes a potential solution and its energy reflects the objective function $H_{\mathrm{Ising}}$, as
\begin{equation}
\min \quad H_{\mathrm{Ising}}(\mathbf{s})
=
\sum_{i=1}^{n} h_i s_i
+
\sum_{i=1}^{n-1}\sum_{j=i+1}^{n} J_{ij}s_i s_j ,
\label{eq:ising_model}
\end{equation}
where \(\mathbf{h}\in\mathbb{R}^{n}\) denotes the linear coefficient vector, \(\mathbf{J}\in\mathbb{R}^{n\times n}\) denotes the upper triangular coupling matrix, and \(s_i\in\{-1,+1\}\) is the spin variable. 
Then, \eqref{eq:ising_model} provides a quantum compatible form, which is equivalent to the QUBO formulation \cite{kochenberger2014unconstrained}.

\subsection{Quantum-Compatible Reformulated QUBO Model}
Since a standard QUBO model does not contain any constraints, the clustering model in \eqref{eq:clustering_obj}–\eqref{eq:assignment_constraint}, which includes equality constraints, must be transformed into a new objective function \(H_{\mathrm{QUBO}} \), expressed as:

\begin{equation}
H_{\mathrm{QUBO}}=
H_{\mathrm{obj}}
+\lambda \sum_{i=1}^{N}
\left(\sum_{g=1}^{G}x_i^g-1\right)^2,
\label{eq:qubo_obj}
\end{equation}
where \(\lambda>0\) is a penalty coefficient.

For compact representation, the resulting QUBO model can be written in matrix form as 
\begin{equation}
H_{\mathrm{QUBO}}=\mathbf{x}^{\top}\mathbf{Q}\mathbf{x},
\label{eq:qubo_standard}
\end{equation}
where \(\mathbf{x}\) denotes the binary decision vector and \(\mathbf{Q}\) is the corresponding coefficient matrix.

Given this QUBO model, the renewable energy profile clustering problem can be solved directly on quantum hardware.

\begin{remark}
Using this QUBO model, the clustering problem becomes compatible with quantum hardware. However, this reformulation is typically constrained by the limited number of available qubits.  In \eqref{eq:qubo_standard}, the number of binary variables scales as \(N \times G\), indicating growth with both the number of profiles \(N\) and the number of clusters \(G\). For QUBO-based quantum hardware, each binary variable is typically mapped to one physical qubit. Consequently, realistic renewable energy clustering tasks can quickly exceed the practical capacity of current quantum hardware, making direct implementation challenging. This motivates us to develop a more qubit-efficient strategy that is described next.
\end{remark}

\section{Proposed Method}
In this section, we first introduce variable reduction based on the one-hot constraints. We then use consensus analysis to identify the stable co-assignment relation. Using it, we can convert them into additional equality constraints for further variable reduction. Finally, the overall CBVR procedure is summarized.

\subsection{Variable Reduction Based on the One-Hot Constraints}

To reduce the number of variables in the QUBO model, the one-hot constraints in \eqref{eq:assignment_constraint} are first applied to remove one binary variable from each renewable profile assignment. Specifically, for each profile \(i\), \(x_i^{G}\) is selected as the dependent variable, while \(\{x_i^1,x_i^2,\dots,x_i^{G-1}\}\) are retained as independent variables. Then, using the variable relationship implied by the one-hot constraint \eqref{eq:assignment_constraint}, we have

\begin{equation}
x_i^{G}=1-\sum_{g=1}^{G-1} x_i^g.
\label{eq:dependent_var}
\end{equation}

Substituting \eqref{eq:dependent_var} into \eqref{eq:clustering_obj} yields
\begin{equation}
\begin{aligned}
H_{\mathrm{obj}}
=&
\sum_{i=1}^{N}\sum_{j=1}^{N} w_{i,j}
\Bigg[
\sum_{g=1}^{G-1} x_i^g x_j^g \\
&\quad+
\left(
1-\sum_{g=1}^{G-1} x_i^g
\right)
\left(
1-\sum_{h=1}^{G-1} x_j^h
\right)
\Bigg].
\end{aligned}
\label{eq:reduced_obj}
\end{equation}

Accordingly, the penalty term induced by the one-hot constraints is then written as \(H_{\mathrm{bin}}\), expressed as

\begin{equation}
H_{\mathrm{bin}}
=
\lambda \sum_{i=1}^{N} x_i^{G}(x_i^{G}-1)
=
2\lambda \sum_{i=1}^{N}\sum_{1\le g<h\le G-1} x_i^g x_i^h.
\label{eq:svr_penalty}
\end{equation}

Thus, the resulting QUBO model is
\begin{equation}
\label{eq:hsvr}
H_{\mathrm{QUBO}}=H_{\mathrm{obj}}+H_{\mathrm{bin}}
\end{equation}

This transformation reduces the dimensionality of the QUBO model from \(NG\) to \(N(G-1)\). However, the reduction is limited to removing only a single variable per profile, resulting in a linear decrease with respect to the number of clusters. As a consequence, the overall problem size remains large when \(N\) or \(G\) is significant. To achieve further variable reduction, a co-assignment scheme to construct extra equality constraints is elaborated next.

\subsection{Variable Reduction Based on Consensus Analysis}

Further variable reduction cannot rely solely on the equality relations induced by the one-hot constraints. Instead, additional equality constraints should be derived from stable pairwise assignment patterns in candidate clustering solutions. Therefore, multiple annealing runs are performed to generate candidate partitions, and consensus analysis is employed to identify profile pairs that are repeatedly assigned to the same cluster in these solutions \cite{Wilkerson2010ConsensusClusterPlus}. Such co-assignment relations can then be converted into additional equality constraints for further variable reduction.

  Specifically, we perform R annealing runs to generate a set of feasible solutions for the model \eqref{eq:hsvr}. For the \(r\)-th annealing run, let \(\hat x_i^{g,(r)}\) denote the variable value, where \(\hat x_i^{g,(r)}=1\) means that profile \(i\) is assigned to cluster \(g\) in run \(r\); otherwise,  \(\hat x_i^{g,(r)}=0\). The reduced assignment variable is recovered as
\begin{equation}
\hat{x}_i^{G,(r)}
=
1-\sum_{g=1}^{G-1}\hat{x}_i^{g,(r)} .
\end{equation}
If \(\hat x_i^{g,(r)}=\hat x_j^{g,(r)}=1\) for some g , they are regarded as being co-assigned in that run. Then, the co-assignment value between profiles \(i\) and \(j\) is defined as
\begin{equation}
\label{eq:coassociation}
C_{i,j}
=
\frac{1}{R}
\sum_{r=1}^{R}
\sum_{g=1}^{G}
\hat x_i^{g,(r)} \hat x_j^{g,(r)} .
\end{equation}

Here, \(C_{i, j}\) measures how consistently profiles \(i\) and \(j\) are grouped together across candidate solutions.

To identify the most stable co-assignment relations while controlling the number of additional equality constraints, all distinct profile pairs \((i,j)\), \(i<j\), are ranked in descending order of their co-assignment values. A larger \(C_{ij}\) indicates that profiles \(i\) and \(j\) are more consistently assigned to the same cluster across multiple annealing runs, and therefore represent a more reliable co-assignment relation. The top-\(K\) ranked pairs are then selected to form the set
\begin{equation}
\mathcal{E}
=
\left\{
(i,j)\mid i<j,\,
\operatorname{rank}(C_{i,j})\le K
\right\}.
\end{equation}
where \(K\) denotes the number of selected profile pairs, and \(\operatorname{rank}(C_{i,j})\) denotes the descending order rank of \(C_{i,j}\).

For each selected pair \((i,j)\in\mathcal{E}\), the corresponding decision variables are tied together as
\begin{equation}
x_i^g=x_j^g,
\qquad
\forall (i,j)\in\mathcal{E},\;
\forall g=1,\dots,G-1.
\label{eq:stable_equality}
\end{equation}
This allows the paired profiles to share a common variable representation. If multiple equality relations contain common profiles, they are further grouped into a larger co-assignment profile set. 

The selected reliable co-assignment pairs in \(\mathcal{E}\) are first aggregated into a set of merged profile groups, denoted by \(\mathcal{M}\), where \(|\mathcal{M}|=M\). Profiles that are not part of any merged group remain as individual profiles and are collected in the set \(\mathcal{U}\), where \(|\mathcal{U}|=U\). Together, the merged groups and the unmerged profiles constitute the updated profile set
\begin{equation}
\mathcal{S}
=
\mathcal{M}\cup\mathcal{U}
=
\{S_s\}_{s=1}^{N'},
\end{equation}
where \(N'=M+U\) denotes the total number of profiles after this variable reduction procedure.
Each element \(S_s\in\mathcal{S}\) is subsequently regarded as a \emph{super profile}, which represents all original profiles contained in the corresponding merged group. The size of each super profile is defined as
\begin{equation}
\kappa_s = |S_s|,
\qquad s=1,\dots,N',
\label{eq:superprofile_size}
\end{equation}
where \(\kappa_s>1\) indicates that \(S_s\) contains multiple original profiles, while \(\kappa_s=1\) indicates that \(S_s\) contains only one original profile.

  For each element \(S_s\) in the updated profile set \(\mathcal{S}\), a representative decision variable \(z_s^g\) is introduced. All original variables associated with profiles in the same group are unified through
\begin{equation}
x_i^g = z_s^g,
\qquad
\forall i\in S_s,\ \forall g=1,\dots,G-1.
\label{eq:representative_variable}
\end{equation}

Substituting \eqref{eq:representative_variable} into \eqref{eq:hsvr}, the QUBO model \eqref{eq:hsvr} can be reformulated into \({H}^{'} \), expressed as

\begin{equation} 
\label{eq:aggregated_qubo} 
\begin{aligned} 
{H}^{'}= 
&\sum_{s=1}^{N'}\sum_{t=1}^{N'}
\bar{w}_{s,t}
\Biggl[
\sum_{g=1}^{G-1} z_s^g z_t^g
+
\Bigl(1-\sum_{g=1}^{G-1} z_s^g\Bigr)
\Bigl(1-\sum_{h=1}^{G-1} z_t^h\Bigr)
\Biggr] \\ 
&\quad
+2\lambda 
\sum_{s=1}^{N'} 
\kappa_s
\sum_{1\le g<h\le G-1} 
z_s^g z_s^h .
\end{aligned} 
\end{equation} 
 
Here, \(\bar{w}_{s,t}\) denotes the aggregated dissimilarity coefficient between super profiles \(S_s\) and \(S_t\), defined as 
 
\begin{equation} 
\label{eq:aggregated_weight} 
\bar{w}_{s,t}
=
\sum_{i\in S_s}
\sum_{j\in S_t} 
w_{i,j}.
\end{equation}

As a result, the proposed method compresses the QUBO model from \(NG\) to \(N'(G-1)\) binary variables. The overall workflow for implementing the proposed CBVR method is summarized in \textbf{Algorithm 1}.





 

\begin{algorithm}[t]

\caption{CBVR Method for Renewable Energy Profile Clustering}
\label{alg:proposed_qubo_reduction}

\begin{algorithmic}[1]
\Procedure{CBVR}{}

\State Compute \(w_{i,j}\) and construct \(H_{\mathrm{QUBO}}\)
by \eqref{eq:qubo_obj}

\For{\(i=1,\ldots,N\)}
    \State \(x_i^{G}\gets
    1-\displaystyle\sum_{g=1}^{G-1}x_i^g\)
\EndFor

\State Construct the reduced QUBO by \eqref{eq:hsvr}

\For{\(r=1,\ldots,R\)}
    \State \(\hat{\mathbf{x}}^{(r)}
    \gets\operatorname{Anneal}(H_{\mathrm{QUBO}})\)

    \Comment{Using a classical computer}

    \State \(\hat{x}_i^{G,(r)}
    \gets1-\displaystyle\sum_{g=1}^{G-1}
    \hat{x}_i^{g,(r)},\ \forall i\)
\EndFor

\For{\(i=1,\ldots,N-1\)}
    \For{\(j=i+1,\ldots,N\)}
        \State \(C_{i,j}\gets
        \displaystyle\frac{1}{R}
        \sum_{r=1}^{R}\sum_{g=1}^{G}
        \hat{x}_i^{g,(r)}
        \hat{x}_j^{g,(r)}\)
    \EndFor
\EndFor

\State \(\mathcal{E}\gets
\operatorname{TopK}
\bigl(\{C_{i,j}\mid i<j\},K\bigr)\)

\State Impose \(x_i^g=x_j^g,\ 
\forall(i,j)\in\mathcal{E},\
g=1,\ldots,G-1\)

\State \(\mathcal{S}\gets
\operatorname{MergeProfiles}(\mathcal{E})
=\{S_s\}_{s=1}^{N'}\)

\For{\(s=1,\ldots,N'\)}
    \State \(\kappa_s\gets|S_s|,\quad
    x_i^g\gets z_s^g,\ \forall i\in S_s\)

    \For{\(t=1,\ldots,N'\)}
        \State \(\bar{w}_{s,t}\gets
        \displaystyle\sum_{i\in S_s}
        \sum_{j\in S_t}w_{i,j}\)
    \EndFor
\EndFor

\State Construct \(H'\) by \eqref{eq:aggregated_qubo}

\State \(\mathbf{z}^{*}\gets
\operatorname{Optimize}(H')\)
\Comment{Using an Ising machine}

\State Recover the original profile labels from
\(\mathbf{z}^{*}\) and \(\mathcal{S}\)

\State \Return cluster labels

\EndProcedure
\end{algorithmic}
\end{algorithm}

\section{Case Study}

Now, let us evaluate the proposed method on the real CIM with 550 qubits provided by QBoson \cite{kaiwu_community}.   Wind turbine power profiles are obtained from the real world Spatial Dynamic Wind Power Forecasting (SDWPF) dataset \cite{zhou2024sdwpf}, which contains SCADA measurements collected at 10-min intervals from a wind farm operated by China Longyuan Power Group Corp. Ltd. in China. The active power measurements are subsequently averaged to an hourly resolution to construct daily wind power profiles, resulting in 24 data points per day. Profiles are normalized to the range \([0,1]\) as
\begin{equation}
\tilde{p}_{i,t}
=
\frac{p_{i,t}-p_{\min}}
{p_{\max}-p_{\min}},
\label{eq:data_normalization}
\end{equation}
where \(p_{i,t}\) denotes the active power of profile \(i\) at time \(t\), and \(p_{\min}\) and \(p_{\max}\) denote the minimum and maximum active power values in the retained dataset, respectively. A total of 300 valid profiles are randomly selected from the preprocessed dataset for the following experiments.

Simulated Quantum Annealing (SQA), random subQUBO (RSQ), Gurobi 11.0, and spin-variable reduction (SVR) are compared on a classical computer.  The parameter settings of RSQ follow those adopted in \cite{Atobe2022}, with the subQUBO size fixed at 50 binary variables. 

Two groups of experiments are considered.
The first group compares the clustering methods across three test cases,
in which 300 renewable profiles are clustered into $2$, $3$, and $4$ groups,
resulting in original QUBO models with $300\times2=600$,
$300\times3=900$, and $300\times4=1200$ binary variables, respectively.
The second set evaluates the proposed method on datasets containing
$N = 150$, $200$, $250$, and $300$ renewable profiles,
with the number of groups fixed at $G = 4$,
corresponding to original QUBO models with $150\times4=600$,
$200\times4=800$, $250\times4=1000$ and $300 \times 4 = 1200$ binary variables, respectively.
The penalty coefficient $\lambda$ is set to fifteen times the maximum
entry of the coefficient matrix \cite{Kumar2018}.  Clustering performance is evaluated using the silhouette coefficient (SC) \cite{Rousseeuw1987}, Davies--Bouldin index (DBI) \cite{Davies1979}, and Calinski--Harabasz index (CHI) \cite{Calinski1974}.

\begin{figure}[!t]
\centering
\includegraphics[
    width=\columnwidth,
    height=0.9\textheight,
    keepaspectratio
]{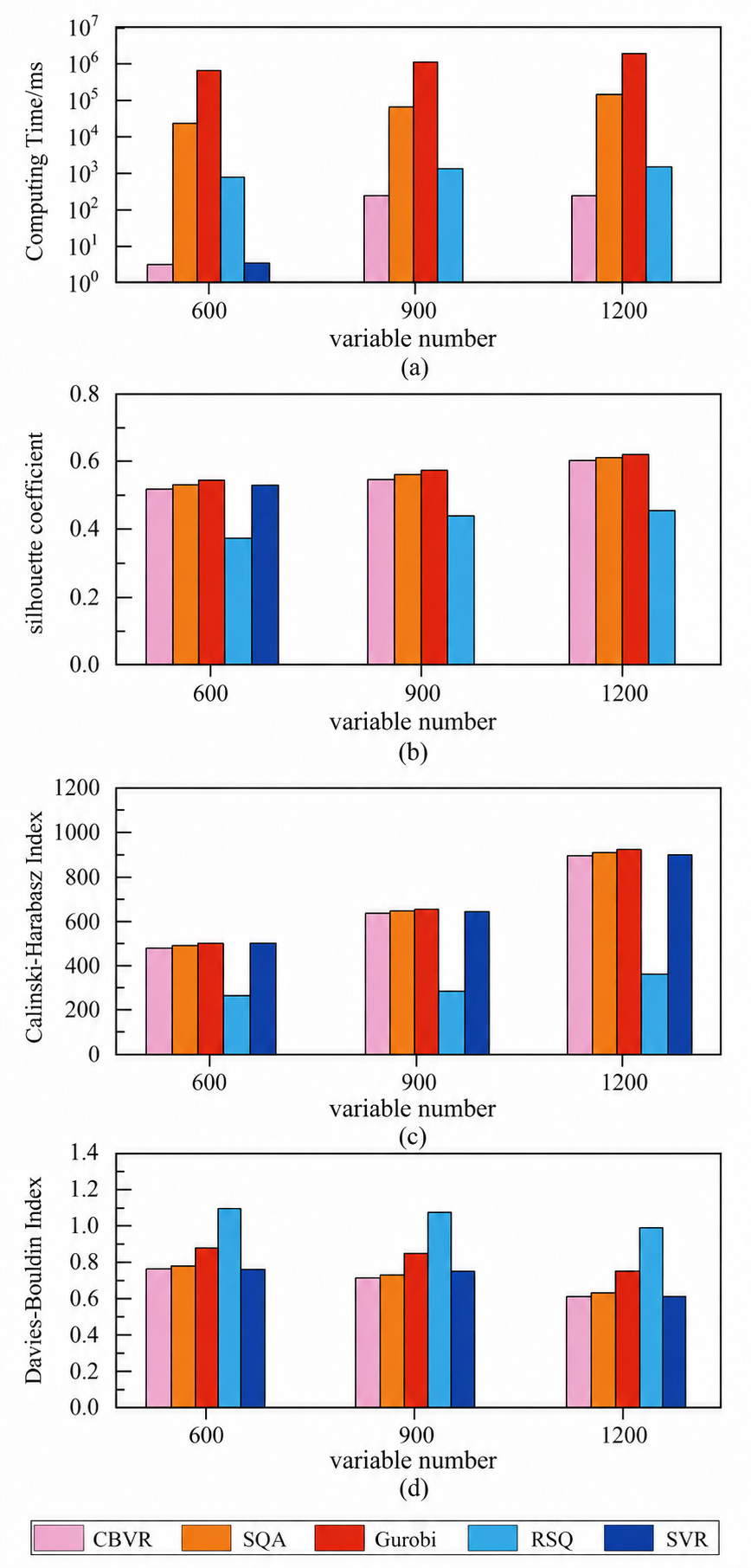}
\caption{Clustering results of different algorithms under different QUBO variable sizes:
(a) computing time;
(b) silhouette coefficient;
(c) Calinski--Harabasz index;
(d) Davies--Bouldin index.}
\label{fig:clustering_results}
\end{figure}

\subsection{Comparison of Clustering Methods}
In the first experiment, since all original QUBO models exceed the 550-variable limit, a variable reduction is required prior to hardware execution. For the  \(300\times2\) case, SVR based only on the one-hot constraints reduces the QUBO size to 300 variables.  For the CBVR preprocessing stage, \(R=100\) preliminary annealing runs are conducted, with each run assigned a nominal time budget of \(T_{\mathrm{a}}=1\,\mathrm{s}\) to generate a candidate clustering solution for consensus analysis. Each run employs a simulated annealing (SA) sampler that repeatedly performs annealing with one sample and 200 sweeps until the prescribed time budget is reached \cite{kaiwu_community}. For the \(300\times3\) and \(300\times4\) cases, the resulting co-assignment relations are ranked according to their co-assignment values, and the top 25 and 117 relations are selected, respectively, to satisfy the 550-qubit hardware limit. This reduces the QUBO sizes to 275×2=550 and 183×3=549, respectively, both within the hardware constraint. 
\subsubsection{Computational Performance}
The computing times of different methods are compared in Fig.~\ref{fig:clustering_results}(a). SQA and Gurobi are applied directly to the original \(300\times2\), \(300\times3\), and \(300\times4\) QUBO models, and therefore their runtimes increase markedly as the variable dimension grows. RSQ restricts each subproblem to 50 variables and achieves shorter runtime, but this acceleration is obtained at the expense of clustering accuracy. Although CBVR involves both quantum hardware execution time and classical preprocessing time for identifying stable co-assignment relations, the total overhead remains substantially lower than the optimization time required by SQA and Gurobi. Moreover, the proposed method achieves a favorable trade-off between computational efficiency and accuracy.

\subsubsection{Clustering Effectiveness}
The clustering performance under different QUBO variable scales is evaluated using the silhouette coefficient, CHI, and DBI, as shown in Fig.~\ref{fig:clustering_results}(b)-(d). These complementary metrics jointly evaluate how closely samples are grouped within each cluster and how well different clusters are separated, where larger silhouette coefficient and CHI values and smaller DBI values indicate better clustering performance. Overall, CBVR exhibits clustering performance comparable to SQA and Gurobi across all three cases. In the \(300\times2\) case, CBVR, SQA, Gurobi, and SVR show similar clustering quality, whereas RSQ exhibits a lower silhouette coefficient and CHI together with a higher DBI. Similar trends are observed in the \(300\times3\) and \(300\times4\) cases, where CBVR remains close to SQA and Gurobi across all three metrics, while RSQ consistently produces less compact and less separated clusters. The consistent behavior across the three validation metrics indicates that CBVR preserves the underlying clustering structure after variable reduction while enabling the resulting QUBO model to satisfy the hardware-size constraint.

The QUBO sizes in different methods are compared in Table~\ref{tab:qubo_size_comparison}. SVR reduces the \(300\times2\) case to 300 variables through assignment-induced equality elimination, whereas it is insufficient to make the \(300\times3\) and \(300\times4\) cases compatible with the quantum hardware limit. By contrast, CBVR reduces the three cases to 300, 550, and 549 variables, respectively, thereby enabling all instances to be executed on the quantum hardware. These results demonstrate that the proposed method effectively compresses the original QUBO model and significantly improves hardware feasibility for larger clustering tasks.
\begin{table}[t]
\centering
\caption{QUBO Variable Size Used by Different Methods}
\label{tab:qubo_size_comparison}
\begin{tabular}{c|ccccc}
\hline
\textbf{Case} & \textbf{CBVR}& \textbf{SQA} & \textbf{Gurobi} & \textbf{RSQ} & \textbf{SVR} \\
\hline
1 & 300 & 600  & 600  & 50 & 300 \\
2 & 550 & 900  & 900  & 50 & --  \\
3 & 549 & 1200 & 1200 & 50 & --  \\
\hline
\end{tabular}
\end{table}

\subsection{Effect of Dataset Size}
In the second experiment, variable reduction is required prior to hardware execution. For the \(150\times4\) case, SVR based only on the one-hot constraints reduces the QUBO size to 450 variables. For the \(200\times4\), \(250\times4\), and \(300\times4\) cases, the one-hot  constraints first reduce the QUBO sizes to 600, 750, and 900 variables, respectively. For the CBVR preprocessing stage, the preliminary annealing parameters and SA sampler settings are the same as those in the first set of experiments. The resulting co-assignment relations are ranked according to their co-assignment values, and the top 17, 67, and 117 relations are selected, respectively, to satisfy the  550-qubit hardware limit. This further reduces all three QUBO models to $183 \times 3 = 549$ variables, each within the hardware constraint.
\subsubsection{Clustering Effectiveness}

The clustering performance under different dataset sizes
is reported in Table~\ref{tab:dataset_size}.
Overall, as the number of profiles increases from 150 to 300,
the silhouette coefficient and DBI exhibit limited variation,
remaining within 0.588--0.622 and 0.688--0.711, respectively,
while CHI shows an increasing trend.
The larger cases requiring additional consensus analysis
achieve silhouette coefficient and DBI values comparable
to those of the smaller case using only SVR,
indicating similar clustering compactness and separation
across the tested dataset sizes.
Together with the variable reduction results, these findings
support the effectiveness of CBVR in accommodating
larger datasets within the same 550-variable hardware limit
while achieving clustering quality comparable to that
of the smaller dataset.

\begin{table}[t]
\centering
\caption{Clustering Results for Different Dataset Sizes
with $G=4$}
\label{tab:dataset_size}
\footnotesize
\setlength{\tabcolsep}{10pt}
\renewcommand{\arraystretch}{1.15}
\begin{tabular}{c|ccc}
\hline
$N$ & \textbf{SC} & \textbf{CHI} & \textbf{DBI} \\
\hline
150 & 0.622 & 466.776 & 0.688 \\
200 & 0.605 & 537.230 & 0.691 \\
250 & 0.588 & 637.765 & 0.711 \\
300 & 0.612 & 814.935 & 0.693 \\
\hline
\end{tabular}
\end{table}

\section{Conclusion and future work}
This paper addresses the problem of clustering renewable energy profiles under limited qubits of current quantum hardware. Hence, a CBVR method is proposed. Experimental results demonstrate that the proposed method can reduce oversized QUBO models to hardware-admissible formulations while maintaining high computing accuracy and efficiency for renewable energy profile clustering. 

Here the variable reduction performance of CBVR depends on the reliability of the co-assignment relations identified from preliminary annealing solutions, which may vary with the problem characteristics and annealing quality. Future work will investigate adaptive strategies for selecting reliable co-assignment relations and evaluate CBVR on larger-scale renewable energy datasets and quantum hardware. Moreover, in the future, bad data or missing data handling strategy will be further considered to improve the robustness of the proposed method.


\FloatBarrier
 \newcommand{\BIBdecl}{\setlength{\itemsep}{0.02 em}}
 \bibliographystyle{IEEEtran}
\bibliography{IEEEabrv,References.bib}

\end{document}